\documentclass[aps,prb,twocolumn,superscriptaddress]{revtex4}
\usepackage{graphicx}
\usepackage{epstopdf}
\usepackage{latexsym}
\usepackage{amssymb}
\usepackage{amsmath}
\usepackage{amsfonts}
\usepackage{subfigure}
\usepackage{bm}
\usepackage{multirow}
\usepackage{color}

\renewcommand{\Re}{\mathop{\mathrm{Re}}}
\renewcommand{\Im}{\mathop{\mathrm{Im}}}
\newcommand{\vect}[1]{{\bm{#1}}}
\newcommand{\eqnref}[1]{Eq.\,\eqref{#1}}
\newcommand{\figref}[1]{Fig.\,\ref{#1}}
\newcommand{\tabref}[1]{Tab.\,\ref{#1}}
\newcommand{\diag}[2]{\vcenter{\hbox{\includegraphics[height=#2]{#1}}}}

\begin{document}
\title{Coexisting itinerant and localized electrons}
\author{Yi-Zhuang You}
\address{Department of Physics, University of California at Santa Barbara}
\author{Zheng-Yu Weng}
\address{Institute for Advanced Study, Tsinghua University, Beijing, 100084}
\date{\today }

\begin{abstract}
The surprising discovery of high-$T_c$ superconductivity in iron-based compounds has prompted an intensive investigation on the role of interaction and magnetism in the these materials. Based on the general features of multi-bands and intermediate coupling strengths, a phenomenological theory of coexisting itinerant and localized electrons has been proposed to describe the low-energy physics in iron-based superconductors. It provides a unified framework to understand magnetic, superconducting, and normal phases, subject to further microscopic justification and experimental verification.
\end{abstract}

\maketitle
\tableofcontents

\section{Introduction}

In the past five years, the study of iron-based superconductors have attract much attention in the condensed matter community and beyond. These new superconductors contain the FeAs/FeSe active layers, and the first principle calculations\cite{Haule:2008dq, Ma:2008ay, Kuroki:2008qq, Singh:2008pt, Subedi:2008fv} have shown that the electrons from the iron $3d$ orbitals dominate the density of states at the Fermi energy.

One key issue under early debate is about whether the iron $3d$ electrons should be treated as itinerant electrons or local moments. Underlying this dispute are the two different schools of thought about the mechanism for superconductivity: the Bardeen-Cooper-Schieffer (BCS) theory in a weakly interacting metal\cite{Bardeen:1957vn,Bardeen:1957kx} versus the resonating valence bound (RVB) type of theory for a strongly correlated system like the cuprate.\cite{Anderson:1987fk,Anderson:2004dk}

The BCS theory is based on a weakly correlated Fermi liquid state of itinerant electrons. At low temperatures, a Fermi liquid state will become unstable against any weak attractive interaction, which drives the electrons near Fermi surface to form Cooper pairs and condense, giving rise to superconductivity. The effective attraction can be either mediated by phonons\cite{Bardeen:1955cr}, plasomons\cite{Kohn:1965qf}, magnons\cite{Berk:1966bh, Doniach:1966dq} etc., or originated from the bare electron interaction via the fluctuation-exchange\cite{Bickers:1989kl} (FLEX) or the renormalization group\cite{Shankar:1994tg} (RG) approaches. To make the attraction dominant at low-energy, the Coulomb repulsion must be effectively screened, which in turn requires the electrons be itinerant, such that the BCS theory usually works for a system of metallic normal state.

On the other hand, in a strongly correlated system like the cuprate,\cite{Anderson:1987fk,Anderson:2004dk} the electrons can be localized, due to the strong Coulomb interaction, to form a Mott insulator at half-filling. Here its charge degree of freedom is gapped, while the remaining spin degree of freedom forms a lattice of fluctuating local moments. Superconductivity arises upon charge doping into the Mott insulator. The failure of a conventional BCS theory lies in that the Coulomb repulsion becomes a dominant effect in shaping the electronic state instead of simply getting screened in the BCS theory. In such a single-band strongly correlated system, electron fractionalization is a natural consequence in which doped charges and localized spins behave distinctly as separated degrees of freedom. The underlying mechanism for superconductivity is generally known as an RVB theory because the singlet pairing of the local moments becomes partially charged upon doping, resembling the Cooper pairing,\cite{Anderson:1987fk} to give rise high-$T_c$ superconductivity.

While the BCS theory and the RVB theory lie in the opposite extremes of the electron correlations, most of studies seem to agree on that the iron-based compound is an intermediate correlated electron system.\cite{Qazilbash:2009ly, Johnston:2010yj} In particular, the multi-band iron $3d$ electrons are involved in the low-energy sector in contrast to the single-band $3d$ electrons in the cuprate. As to be discussed in this Chapter, a new possibility\cite{Kou:2009bd} may arise in the low energy regime of the iron-based superconductor, in which the multi-band $3d$ electrons effectively behave as if some of them still remain itinerant near the Fermi energy and some of them become more localized like in a Mott insulator. Such a ``fractionalization'' into two more \textit{conventional} subsystems of itinerant and localized electrons in a multi-band case is in sharp and interesting contrast with a \textit{novel} fractionalization of electrons in a single band doped Mott insulator. Here, without doping into the Mott insulator, the itinerant electrons remain effectively separated from the local moments  as independent degrees of freedom, and two subsystems mutually interact with an intermedate coupling strength, which can be tractable perturbatively. In the following, relevant experimental facts and theoretical approaches will be briefly overviewed.

\subsection{Basic Experimental Evidence}

The experimental evidence for the simultaneous presence of both itinerant electrons and local moments has been manifested in almost all families of iron-based superconductors.

\subsubsection{Itinerant Electrons}

\begin{figure}[htbp]
\begin{center}
\includegraphics[width=140pt]{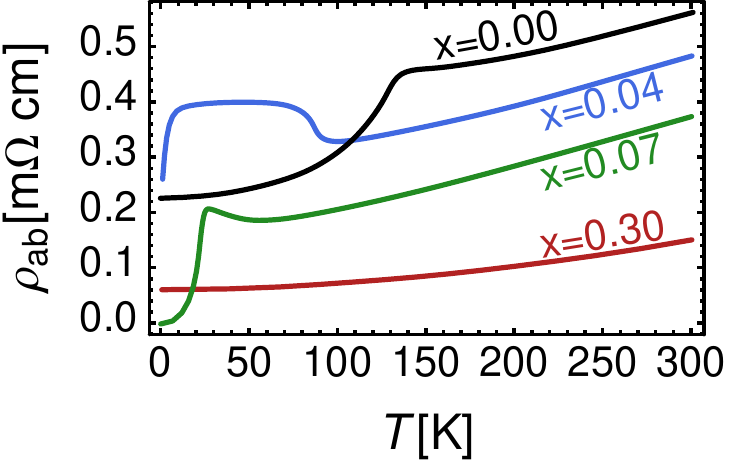}
\caption{The in-plane resistivity $\rho_{ab}$ of Ba(Fe$_{1-x}$Co$_x$)$_2$As$_2$ v.s. the temperature $T$. The curves show the cases at various doping levels, including the parent compound $x=0$, under-doped $x=0.04$, optimal-doped $x=0.07$ and over-doped $x=0.30$. Cited from Ref.\,\onlinecite{Fang:2009dq}.}
\label{fig: resistivity}
\end{center}
\end{figure}

An early direct experimental fact that supports the existence of itinerant electrons in the iron-based compounds is the semi-metal behavior even in the magnetically ordered phase. For most families of the materials,\cite{Kamihara:2008jo,Ren:2008vn,Ren:2008ys,Chen:2008kx,Kito:2008mz,Tapp:2008lh,Sefat:2008qa,Rotter:2008kl,Wang:2008ff,Wang:2008ve,Sales:2009il,Chen:2009fu,Chu:2009fc} it has been observed that the resistivity decreases as the temperature is lowered, as shown in \figref{fig: resistivity} (taking the 122-family\cite{Fang:2009dq} as an example), a typical behavior of a metallic system, contrary to the insulating and localization behavior generically observed in the undoped and lightly doped cuprates in the magnetically ordered phase.
A finite residual resistivity of the parent compound, of the order $\rho_{ab}\simeq0.1$\,m$\Omega\,$cm at low $T$ limit, is shown in \figref{fig: resistivity} (the same order of resistivity is also observed in the 111-family\cite{Tapp:2008lh} and the 11-family\cite{Sales:2009il}). According to the analysis in Refs.\,\onlinecite{Si:2008xi, Johnston:2010yj}, for a two-dimensional electron system, the in-plane resistivity $\rho_{ab}$ is related to the electron mean-free-path $l$ by $k_\text{F}l=(h/e^2)(a_c/\rho_{ab})$, where $k_\text{F}$ is the Fermi momentum, and $a_c$ is the distance between the conducting layers (which is $a_c\simeq 6.5$\AA for the 122-family). According to this estimate, the in-plane electron mean-free-path can be as long as $k_\text{F} l \simeq 14\gg 1$,\cite{Johnston:2010yj} indicating a nice coherence of the quasiparticles around the Fermi surface, which can be identified as well-defined itinerant electrons.

The band structure obtained by the first principle calculations\cite{Singh:2008xq,Cao:2008bh,Ma:2008ay,Kuroki:2008qq} shows that all the five iron $d$ orbitals are close to each other in energy, and the density of states near the Fermi energy are mainly contributed by the Fe $d$ orbitals. The electron and hole Fermi pockets predicted in the band structure calculation are clearly confirmed by angular resolved photoemmision spectroscopy (ARPES) experiments.\cite{Liu:2008yo} Although a renormalization factor of $m^*/m\simeq 2\sim4$ is usually required\cite{Lu:2009lp,Cui:2012kl} to account for the experimental data, which may be due to correlation effect, the fact that there are itinerant electron bands going across the Fermi level is well established. Moreover, the observed broadening of the energy spectrum gets progressively reduced approaching the Fermi level,\cite{Liu:2008yo, Lu:2009lp, Cui:2012kl, Vilmercati:2009rq, Xia:2009dz, Borisenko:2010qo, He:2010bf} which implies asymptotically better quasiparticle coherence. This asymptotic coherence is a typical Landau Fermi-liquid behavior. Similar behavior is also shown\cite{Aichhorn:2009zr} in the dynamic mean-field theory (DMFT) calculations. 

The quantum oscillation observed in the 1111-family\cite{Coldea:2008nx} and 122-family\cite{Sebastian:2008hc,Coldea:2009oq,Analytis:2009kl,Analytis:2009tg} lends further support to the itinerant electron coherency around the Fermi surface. The measurement\cite{Coldea:2009oq} also shows that the mean-free-path of the itinerant electrons can reach the order of 1000\AA, which is much larger than the lattice constant ($\sim 4$\AA). Even in the spin-density-wave (SDW) ordered magnetic parent compounds, because the SDW state is not fully gapped,\cite{Ran:2009or} quantum oscillation experiment can still detect the residual Fermi surface and confirm the quasi-particle coherence\cite{Analytis:2009tg} in the magnetically ordered state. 

The SDW gap opened up from the reconstruction of the itinerant electrons near the Fermi pockets has been also observed in both the scanning tunneling microscopy (STM) and optical conductivity measurements. An SDW gap of $\sim15$meV is found in the STM experiment\cite{Zhou:2012ia} with the gap bottom deviating from the Fermi level due to imperfect nesting. The optical conductivity experiments\cite{Hu:2008ys, Moon:2010ij, Moon:2010bs,Wang:2012ij} observed a low-energy spectral weight transfer in the SDW transition. However both the STM\cite{Zhou:2012ia} and the optical conductivity\cite{Moon:2010bs} measurements have also indicated an energy gap feature ($\sim0.2$eV) substantially larger than the SDW gap, which is present over a much higher temperature and wider doping regime covering both SDW and superconducting (SC) phases. This latter energy scale may be considered to be a generalized Mott gap which protects some effective local moments, implying the coexistence of both itinerant and localized electrons in the system. In the following, the experimental evidence for the presence of local moments will be briefly discussed.

\subsubsection{Local Moments}

The elastic neutron scattering (ENS) study of the magnetically ordered parent compounds has shown that the ordered moment per Fe atom, $\mu_\text{Fe}$, varies significantly among the materials: $\mu_\text{Fe}\sim0.4\mu_B$ in the 1111-family\cite{Cruz:2008ws, Chen:2008dz, Kimber:2008fv, Qiu:2008fu, Zhao:2008bs, Xiao:2009qa, Xiao:2010mi, Tian:2010kl}, $\mu_\text{Fe}\sim0.8\mu_B$ in the 122-family\cite{Goldman:2008tf, Zhao:2008pi, Kaneko:2008pt, Huang:2008um, Matan:2009ti, Wilson:2009lh, Xiao:2009ff}, and $\mu_\text{Fe}\sim2\mu_B$ in the 11-family\cite{Li:2009zl, Bao:2009bs, Martinelli:2010qo,Liu:2010tw}. One may consider the relatively small magnetic moment (compared to that of the Fe$^{2+}$ ion\cite{Cao:2008bh}) observed in the iron-based compounds as an evidence for the itinerant magnetism. However the opposite opinion argues that not all the $d$-orbital electrons participate in the formation of the local moment, as some of them may remain itinerant around the Fermi surface such that a local moment should not be simply deduced from an Fe$^{2+}$ ion model. Further, the ordered moment is subject to the magnetic fluctuation,\cite{Si:2008xi,Rodriguez:2009zt,Hansmann:2010jl} which is averaged over the observation timescale, and is always smaller than an instant moment. 

An instant (high energy scale) moment at the iron atom can be directly measured by the x-ray emission spectroscopy\cite{Gretarsson:2011fk,Vilmercati:2012jf} (XES), in which the measurement timescale is of $10^{-16}\sim10^{-15}$s, much faster than the timescale $10^{-8}\sim10^{-6}$s of the Mössbauer spectroscopy, the nuclear magnetic resonance (NMR) and the muon spin resonance ($\mu$SR), and is also faster than the ENS timescale by $1\sim2$ orders of magnitude. So the XES measurement probes the instant magnetic moment at a time scale much shorter than that of magnetic correlations established among the local moments as probed in the ENS experiment. The XES result\cite{Gretarsson:2011fk} shows that even at room-temperature, a local magnetic moment can still be detected, which is about $\mu_\text{Fe}\simeq1\mu_B$ (corresponding to the spin $S=1/2$) for the 1111-, 122- and 111-families, and $\mu_\text{Fe}\simeq2\mu_B$ for the 11-family. It is further discovered that the local moment exists in all phases including the magnetically disordered phases as well as the paramagnetic (PM) and SC phases. 

Here the size of the local moment is insensitive to the temperature variation, which excludes the possibility that these moments are originated from itinerant magnetism. A careful experiment study\cite{Vilmercati:2012jf} discovers that the local moment varies in different phases. The measured $\mu_\text{Fe}\simeq1.3\mu_B$ in the PM/SC phases of the optimal doped Sr(Fe$_{1−x}$Co$_x$)$_2$As$_2$($x=0.1$) is reduced by half as compared to $\mu_\text{Fe}\simeq2.1\mu_B$ of the parent compound SrFe$_2$As$_2$ in the SDW phase at 17K, as if the local moment spin is reduced from $S=1$ in the SDW phase to $S=1/2$ in the PM/SC phase, indicating the possibility of spin fractionalization of the local moment outside the SDW phase. 

Other indirect evidences for the existence of local moments include that the nuclear hyperfine splitting in the Mössbauer spectrum\cite{Bonville:2010kx} persists up to $1.5T_\text{SDW}$ (where $T_\text{SDW}$ stands for the SDW transition temperature), and that a well-defined spin-wave spectrum observed in the inelastic neutron scattering (INS) experiments\cite{Zhao:2009hc, Xu:2011uq} extends up to the energy scale of 200meV ($\sim1.8T_\text{SDW}$). The pure itinerant electron picture can hardly account for the high-energy/high-temperature magnetism, when the SDW order ceases to exist. 

Mover, the NMR Knight shift\cite{Ning:2009ss, Imai:2009mw, Ning:2010qq, Michioka:2010fc,Ma:2011ee} and uniform suspectibility\cite{Yan:2008rq, Wu:2008cq, Wang:2009kh, Klingeler:2010mb} experiments have both observed the linear temperature dependence of the magnetic susceptibility all the way to above 500-800K. Such behavior can be explained\cite{Zhang:2009lp,Kou:2009bd} by the antiferromagetic (AFM) short-range correlation between local moments.  Here the experiments once again indicate the persistence of the local moments with AFM correlations up to much higher temperatures than $T_\text{SDW}$.

\subsection{Theories for Iron-Based Superconductors}

In general, there are three schools of theories: itinerant theory\cite{Mazin:2008kn, Yildirim:2008ve, Chubukov:2008bu, Sknepnek:2009km, Maier:2009ez, Wang:2009wq, Zhai:2009nx, Ummarino:2009lg}, localized theory\cite{Si:2008xi, Seo:2008ov, Craco:2008zr, Haule:2009ys, Laad:2009ly, Abrahams:2011uq, Hu:2012ys, Yu:2012kx, Flint:2012pt} and the hybrid theory of coexistent itinerant and localized electrons\cite{Weng:2009if, Kou:2009bd, Yin:2010hs, Lv:2010kc, Gorkov:2013pu}, see \tabref{tab: theory}.

\begin{widetext}
\begin{center}
\begin{table}[htdp]
\caption{Comparison of main-stream theories for iron-based superconductor}
\begin{tabular}{p{3.cm}p{4.2cm}p{4cm}p{4cm}}
\hline
& Itinerant electron & Local moment & Hybrid\\
\hline
Degrees of freedom & Itinerant electrons & Local  moments & Coexistence of both\\
Electron correlation & Weak & Strong & Intermediate \\
Starting point & Fermi liquid & Mott insulator & Orbital-selective Mott  \\
SC mechanism & BCS pairing & Spin RVB pairing & BCS pairing\\
Pairing glue & Electron collective fluctuation  & Superexchange & Local moment fluctuation\\
\hline
\end{tabular}
\label{tab: theory}
\end{table}
\end{center}
\end{widetext}

The itinerant theory is built on the picture of pure itinerant electrons, which views the iron-based superconductor as a simple BCS superconductor with the electron pairing mediated by spin-fluctuations generated by the interaction among itinerant electrons. The local theory takes a strong correlation point of view and considers the iron-based superconductor as a multi-band version of doped Mott insulators similar to the cuprates.  The coexistence theory emphasizes that itinerant electrons and local moments should both exist in the iron-based materia, as independent degrees of freedom at least in the low-energy sector. A careful differentiation of these theories and their underlying physics is important in search for the correct microscopic mechanism of superconductivity, as to be detailed below.

\subsubsection{Itinerant Electron Theory}

Starting from the $3d$-orbital itinerant electron bands, and combining with the intra-atomic interaction, one can establish the multi-band Hubbard model\cite{Raghu:2008pd, Chubukov:2008bu, Lee:2008zp}
\begin{equation}\label{eq: Hubbard}
\begin{split}
H&=H_\text{it}+H_\text{int},\\
H_\text{it}&=\sum_{i,j;\alpha,\beta}t_{ij}^{\alpha\beta} c_{i\alpha}^\dagger c_{j\beta} + h.c.,\\
H_\text{int} &=\frac{1}{2}\sum_{i}\Big(U\sum_{\alpha}n_{i\alpha}n_{i\alpha}+V\sum_{\alpha\neq\beta}n_{i\alpha}n_{i\beta}\\&\hspace{30pt}-J\sum_{\alpha\neq\beta}\vect{S}_{i\alpha}\cdot \vect{S}_{i\beta}+J\sum_{\alpha\neq\beta}\Delta_{i\alpha}^\dagger \Delta_{i\beta}\Big).
\end{split}
\end{equation}
where $c_{i\alpha}=(c_{i\alpha\uparrow},c_{i\alpha\downarrow})^\intercal$ is the electron operator of the $\alpha$ orbital on the $i$ site, which contains two spin components $\uparrow$ and $\downarrow$. Here $n_{i\alpha}=c_{i\alpha}^\dagger c_{i\alpha}$ is the charge density operator, $\vect{S}_{i\alpha}=c_{i\alpha}^\dagger \vect{\sigma} c_{i\alpha}$ is the spin operator, and $\Delta_{i\alpha}=c_{i\alpha}i\sigma_2 c_{i\alpha}$ is the pairing operator. $H_\text{it}$ describes the hopping of electron, in which the hopping coefficient $t_{ij}^{\alpha\beta}$ can be obtained from the band structure calculation\cite{Kuroki:2008qq}, or determined by fitting the ARPES observed band structure. $H_\text{int}$ describes the electron interaction inside the iron atom, including the the intra-orbital repulsion $U$, the inter-orbital repulsion $V$, and the Hunt's rule exchange interaction and the pair-hopping interaction $J$.

Based on the multi-band Hubbard model, an itinerant theory starts from the electron band structure $\epsilon_\vect{k}$ in $H_\text{it}$, and treats the interaction term $H_\text{int}$ perturbatively. The simplest treatment\cite{Sknepnek:2009km, Maier:2009ez} includes the calculation of the spin susceptibility function $\chi=\chi_0(1-\Gamma \chi_0)^{-1}$ in the RPA framework, where $\Gamma$ stands for the interaction vertex given by $H_\text{int}$, while the bare spin susceptibility $\chi_0$ can be calculated from the band structure by $\chi_0(\nu,\vect{q})\simeq -\sum_{\vect{k}}(n_F(\epsilon_{\vect{k}+\vect{q}})-n_F(\epsilon_{\vect{k}}))/(\nu+\epsilon_{\vect{k}+\vect{q}}-\epsilon_{\vect{k}})$ where $n_F$ denotes the Fermi function. The $\chi(q)$ obtained from the RPA calculation reflects the collective spin fluctuation of itinerant electrons under weak interaction. According to the Berk-Schrieffer theory,\cite{Berk:1966bh} the spin fluctuation can mediate pairing interaction between electrons, as $H_\text{pair}\simeq \sum_{k,k'}\Delta_{k}^{\dagger}\chi(k-k')\Delta_{k'}$, where $\Delta_k\sim c_{k}c_{-k}$ is the Cooper pair operator. Plugging this interaction into the Eliashberg gap equation,\cite{Ummarino:2009lg} one can obtain the form factor of $\Delta_k$, and estimate the superconductivity transition temperature. Following this line of thought, Mazin and collaborators\cite{Mazin:2008kn} first predicted the $s_\pm$-wave pairing symmetry in the iron-based superconductor, which is consistent with many experiments. Thus the spin-fluctuation BCS theory\cite{Scalapino:1986os, Pines:1997ee, Pines:1997xw} previously developed for the cuprate superconductors thrives again in the study of  iron-based superconductors. The simple RPA calculation can be improved to a self-consistent FLEX calculation.\cite{Yu:2009vz, Yao:2009rw} Or one can use the RG approaches\cite{Chubukov:2008bu, Chubukov:2009ta, Wang:2009wq, Zhai:2009nx} to track the flow of the interaction vertex towards the low energy scale, so as to analyze the competition between different orders.

Besides the SC phase, the SDW phase may also be understood within the itinerant electron theory. Due to an approximate nesting of the Fermi pockets, the spin fluctuation becomes the strongest near the nesting momentum, which is consistent with the collinear antiferromagnetic (CAFM) order in most parent compounds. In an itinerant electron theory, SDW and SC can compete and coexist.\cite{Vorontsov:2009mq, Vorontsov:2010wa,Fernandes:2010fk} However more concrete calculation by the FLEX method\cite{Arita:2009jb} shows that starting from a purely itinerant picture, it is hard to obtain robust enough CAFM order in a reasonable parameter regime, which implies the importance of local moments in stabilizing the SDW phase. 

\subsubsection{Local Moment Theory}

In view of the bad metal behavior, many consider\cite{Si:2008xi, Craco:2008zr, Haule:2009ys, Laad:2009ly, Abrahams:2011uq, Flint:2012pt} the parent compounds of iron-based superconductors to be proximate to Mott insulators. In other words, the materials are in strongly correlated regime, in contrast to a weakly correlated itinerant electron description outlined above. Based on this idea, a mutiband $t$-$J_1$-$J_2$ model\cite{Si:2009vn} has been proposed:
\begin{equation}
\begin{split}
H=&H_t+H_J,\\
H_t=&\sum_{i,j} t_{ij} \mathcal{P}c_i^\dagger c_j \mathcal{P}+ h.c.,\\
H_J=&\sum_{i,j} J_{ij} \vect{S}_{i}\cdot \vect{S}_{j}.
\end{split}
\end{equation}
where in the electron operator $c_i$, the spin and orbital degrees of freedom are implicitly implied, and $\vect{S}_i=c_i^\dagger \vect{\sigma} c_i$ stands for the local moment at the $i$ site made up by the localized electrons. The superexchange interaction between the local moments is described by $J_{ij}$. The CAMF order in the SDW phase can be reasonably explained by considering the nearest neighboring coupling $J_1$ and the next nearest neighboring coupling $J_2$.\cite{Si:2008xi} $t_{ij}$ describes the hopping of doped electrons on the lattice, and the projection operator $\mathcal{P}$ restricts the on-site electron configuration in the physical Hilbert space to distinguish the local moment and doped charge (which can be regarded as a multi-band generalization of the no-double-occupancy condition in the single-band Hubbard model).

Because of the projection operator $\mathcal{P}$, one can no longer simply treat $H_t$ in Eq. (2) as describing itinerant electrons like in $H_\text{it}$ of Eq. (1). In other words, $\mathcal{P}$ enforces strong correlations in the $t$-$J_1$-$J_2$ model. One way to tackle the $t$-$J_1$-$J_2$ model is to introduce the so-called U(1) slave boson approach.\cite{Flint:2012pt} In a fashion similar to the one-band $t$-$J$ model, based on the picture of spin-charge separation, here the electron operator $c_{i\sigma}$ may be fractionalized into a product of the fermionic spinon $f_{i\sigma}$ and the bosonic chargon $a_{i}$ as $c_{i\sigma}=a_i f_{i\sigma}$, under the equal-density constraint of the spinon and the chargon $a_i^\dagger a_i = \sum_{\sigma} f_{i\sigma}^\dagger f_{i\sigma}$. In the SC phase, the chargons are condensed $\langle a_i\rangle\neq0$, so that at the mean-field level, the spinon dynamics follows $H_\text{MF}=-\sum_{i,j}u_{ij}f_i^\dagger f_j+h.c.+\sum_{i,j}J_{ij}(f_i^\dagger \vect{\sigma} f_i)\cdot (f_j^\dagger \vect{\sigma} f_j)$, where $u_{ij}=t_{ij}\langle a_i^\dagger\rangle\langle a_j\rangle$. For the zero-momentum condensate of chargons, we roughly have $u_{ij}\propto t_{ij}$, meaning that the spinon shares the same band structure as the itinerant electrons in Eq. (1). $H_\text{MF}$ actually describes a spinon Fermi-liquid with magnetic interaction $J_{ij}$. In the momentum space, the magnetic interaction effectively becomes a pairing interaction among the spinons $H_\text{pair}= \sum_{\vect{k},\vect{k}'}\delta_\vect{k}^\dagger J(\vect{k}-\vect{k}')\delta_{\vect{k}'}$, where $J(\vect{q}) = -3\sum_{i,j}J_{ij}e^{i\vect{q}\cdot(\vect{i}-\vect{j})}$, and $\delta_\vect{k}=f_{-\vect{k}}i\sigma_2f_{\vect{k}}$ representes the spin-singlet pairing operator of spinons. The CAFM order of the parent compounds implies that the next nearest neighboring antiferromagnetic exchange interaction $J_2$ is dominant. While in the momentum space, $J_2$ interaction corresponds to $J(\vect{q})=-12J_2\cos q_x \cos q_y$, which is attractive $J(0)=-12J_2$ in the long range, and repulsive $J(\vect{Q}_s)=12J_2$ in the short range (where $\vect{Q}_s=(\pi,0)$ is the nesting momentum between the hole and the electron pockets in the iron-based compounds). This interaction combined with the band structure would naturally give rise to the $s_\pm$-wave pairing symmetry, i.e. the pairing order parameter remains the same sign $\delta_\vect{k}$ within the same pockets so as to gain energy in the $J(0)$ channel, while the pairing sign becomes opposite between the pockets connected by $\vect{Q}_s$ as favored by the $J(\vect{Q}_s)$ channel. Under the chargon condensation, the pairing of the electron $\Delta_\vect{k}=c_\vect{-k}i\sigma_2 c_\vect{k}$ directly follows from the spinon pairing $\Delta_\vect{k} = \langle a\rangle^2 \delta_\vect{k}$, such that the experimentally observed $s_\pm$-wave pairing of the electrons may be similarly understood in the slave-boson theory.

The above analysis indicates a close relation between the magnetic fluctuations of the local moments and the pairing symmetry in iron-based superconductors. The short-ranged CAFM fluctuation, with a momentum $\vect{Q}_s$ that connects the electron and the hole pockets, would always lead to the $s_\pm$-wave pairing symmetry. Thus for both the itinerant electron theory and the local moment theory, the same conclusion on the pairing symmetry can be reached.\cite{Mazin:2009rp, Hu:2012ys} However, just like in the high-$T_c$ cuprate, the pairing symmetry itself is not enough to resolve the mechanism for superconductivity. 

\subsubsection{Hybrid Theory}

The itinerant theory holds the point of view of weak electron correlations, while the local moment theory stresses strong electron correlations where \textit{all} the electrons are in or proximate to the (doped) Mott insulator regime. Two pictures are in opposite limits: i.e., itinerant vs. localized electrons. In the latter case, the metallicity of doped electrons no longer simply behaves like that of itinerant electrons obtained by a band structure calculation because under the projection operator $\mathcal{P}$ in Eq. (2), now the electrons have to always remember that part of them are localized moments. This has been the very key issue in the study of doped Mott insulator, and generally a spin-charge separation or fractionalization of the electron seems to result as the natural consequence of such strong correlations as mentioned above.

However, in contrast to the cuprate superconductors, the iron-based superconductors are the multi-band materials with intermediate coupling strengths, which opens door for a new possibility. Given the experimental facts that the itinerant electrons and local moments are both well established in different channels of measurements as seen above, it is sensible to make a phenomenological hypothesis that may be both degrees of freedom, i.e., itinerant electrons and localized moments, can spontaneously emerge from the 3$d$ electron bands after some intermediate strength interactions in Eq. (1) has been taken into account. Namely, in an RG sense, the following phenomenological model\cite{Weng:2009if, Kou:2009bd, Yin:2010hs} may become relevant to the low-energy physics of the iron-based superconductors
\begin{equation}
\begin{split}
H=& H_\text{it}+H_\text{loc}+H_\text{cp} ,\\
H_\text{it} =& \sum_{i,j} t_{ij} c_i^\dagger c_j + h.c., \\
H_\text{loc} =& \sum_{i,j} J_{ij} \vect{M}_{i}\cdot \vect{M}_{j},\\
H_\text{cp} =& -J_H \sum_{i,j} \vect{M}_{i}\cdot(c_i^\dagger \vect{\sigma} c_i).
\end{split}
\end{equation}
where $H_\text{it}$ captures the itinerant electron band structure which determines the Fermi pockets observed in the ARPES, $H_\text{loc}$ describes the superexchange couplings between the local moments denoted by $\vect{M}_i$, and $H_\text{cp}$ accounts for the simplest residual interaction between the itinerant electrons and the local moments:  a renormalized Hund's rule ferromagnetic coupling $J_H$.

\begin{figure}[htbp]
\begin{center}
\includegraphics[width=200pt]{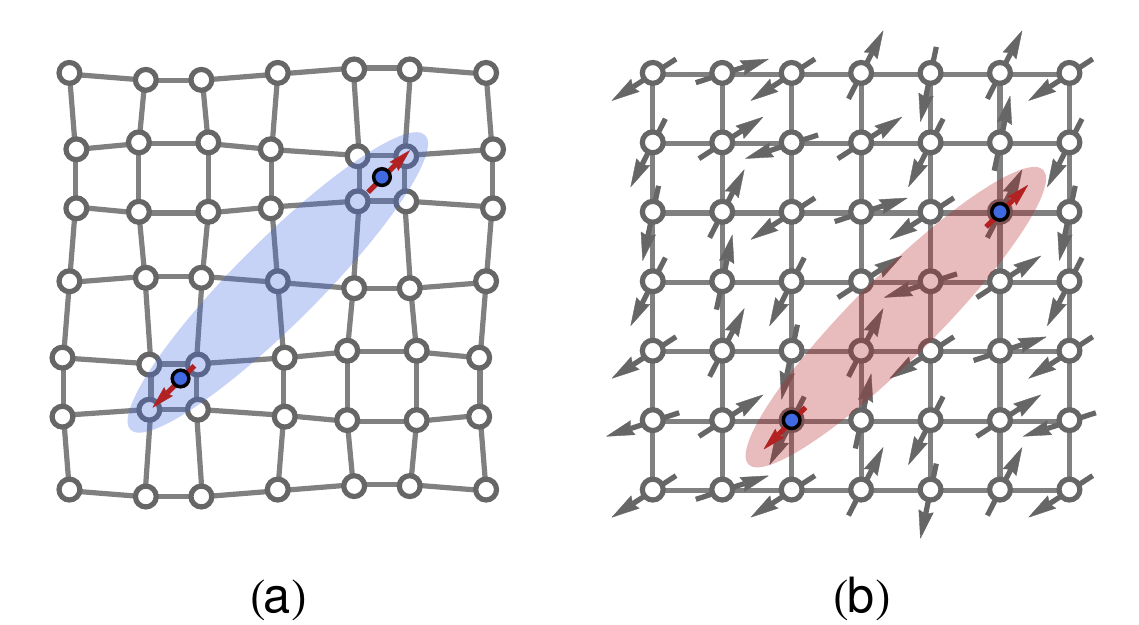}
\caption{The analogy between the phonon-mediated pairing and magnon-mediated pairing: (a) The electron Cooper pairing in a conventional superconductor is glued by the phonons. The phonon couples to the electron charge as the latter induces a lattice distortion which propagates as elastic waves (phonons). (b) The electron Cooper pairing glued by the magnon. Here the itinerant electrons can induce the local moment precession which then propagates as spin waves (magnons).}
\label{fig: BCS}
\end{center}
\end{figure}

If the effective Hund's rule coupling $J_H$ is sufficiently weak, the Hamiltonian (3) simply reduces to two independent states: a Fermi liquid of itinerant electrons and a short-range CAFM state of the local moments, consistent with the observed \textit{normal state} of the iron-based superconductors. The ferromagnetic coupling $J_H$ between the itinerant electrons and the local moments tends to align their spins/magnetic moments in the same direction. If the Fermi surfaces are reasonably well nested, a strong SDW instability will occur to the itinerant electrons by even weakly coupling to the short-ranged CAFM correlation of the local moments. Furthermore, such an SDW order is under a strong competition form the Cooper pairing instability, because by the same coupling term $H_\text{cp}$ the itinerant electrons also tend to pair by exchanging the collective magnon mode of the local moments, as illustrated in \figref{fig: BCS}, in analogy to the conventional phonon-glue BCS superconductor. Because the magnon energy scale ($\sim 100$meV) is much higher than that of conventional phonons, the transition temperature of the magnon-glue BCS superconductivity may exceed the McMillan limit to give rise to a higher $T_c$. 

It is important to distinguish Eq. (3) from Eq. (2) as they represent drastically different low-energy physics. In the effective theory of Eq. (3), the itinerant electron creation operator $c_i^\dagger$ and the local moment operator $\vect{M}_i$ are \textit{independent} degrees of freedom, whereas in Eq. (2) $c_i^\dagger$ and $\vect{S}_i$ are the operators for the \textit{same} electrons. In particular, the strong correlation nature of Eq. (2) is enforced by the Gutzwiller projection operator $\mathcal{P}$. In other words, a simple-minded relaxation of it in Eq. (2) could result in a totally different weak-correlation physics: i.e., an itinerant electron system interacting with perturbative coupling strength, $J_{ij}$, where there is no trace of local moments anymore!

The hybrid theory shares with the itinerant theory that both the SDW and SC orders are formed by the same itinerant electrons. But in the former, the driving force for both SDW instability and the pairing glue comes from coupling to the local moments, whose correlations can persist over a wide temperature and doping regime. In the itinerant theory, however, the SDW instability is driven by the Fermi surface nesting and the pairing glue is attributed to spin fluctuations of the itinerant electrons themselves. Beyond a narrow transition region of the SDW order, such spin fluctuations usually damp quickly, and the Cooper pairing of the itinerant electrons could further suppress the spin fluctuations self-consistently. 

Although the debate on the roles of itinerant electrons and local moments has not been settled completely, it seems that the the consensus is converging to the picture of coexistence in order to account for a vast range of experiements. The introduction of two independent degrees of freedom in the hybrid model is not to complicate the problem, but to separate the different roles played by the iron $d$ electrons, which in turn simplifies the phenomenological description of the iron-based superconductors. Here the itinerant electron and the local moment may be regarded as the emergent degrees of freedom in the multi-band Hubbard model \eqnref{eq: Hubbard} at low energy, resulting from the so-called orbital-selective Mott transition to be discussed below.

\subsubsection{Orbital Selective Mott Transition}

The descriptions of itinerant electrons and local moments are distinguished by the Mott transition. 
The study of Mott transition has a history of more than half a century.\cite{Gutzwiller:1963dq,Hubbard:1963kl} The discovery of the cuprate superconductors has motivated an extensive study of the single-band Hubbard model. It has been demonstrated in the DMFT and quantum Monte-Carlo (QMC) calculations\cite{Georges:1996oq} that there exists an intermediate correlated region where the Mott transition takes place. However, the presence of an SDW/AFM ordering in the itinerant/local moment regime may mask such a transition at low temperatures for a single-band Hubbard model, say, on square lattice.  

On the other hand, because the iron-based superconductor is not only intermediate-correlated but also possesses multiple bands in the electronic structure, a new possibility arises beyond the two simple classifications of itinerant and localized electrons. Here the multi-band structure combined with the intermediate correlation may lead to a new kind of Mott transition: the orbital-selective Mott (OSMott) transition.\cite{Medici:2009qf,Hackl:2009uq,Vojta:2010yq,Medici:2011cr,Yu:2011nx,Zhang:2012bh,Quan:2012dq,Yu:2012ve} With the OSMott transition, different $d$ electron bands will exhibit distinct characteristics of itineracy and Mott localization, which supports the previously outlined hybrid theory of coexisting itinerant electrons and local moments. 

\begin{figure}[htbp]
\begin{center}
\includegraphics[width=150pt]{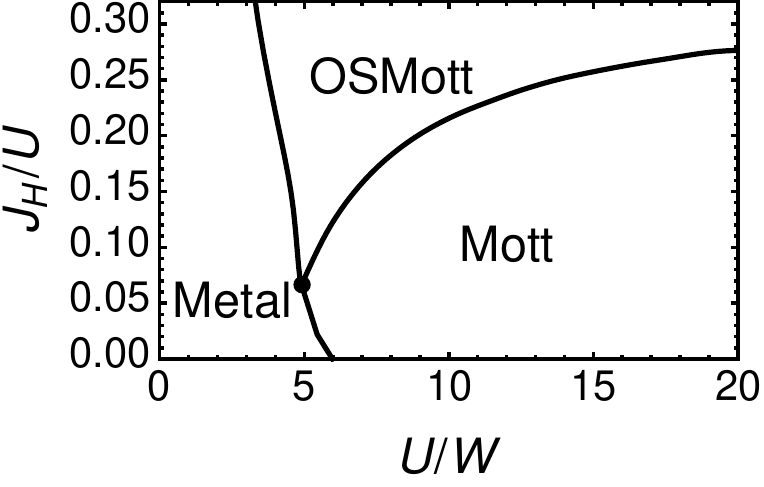}
\caption{The phase diagram of OSMott transition. $U$ is the Hubbard repulsion and $J_H$ is the Hund's rule interaction. $W$ stands for the typical band width. When the Hund's rule coupling reaches certain strength, the OSMott phase emerges in the intermediate correlated region between the metallic and insulating phase. Cited from Ref.\,\onlinecite{Medici:2009qf}.}
\label{fig: OSMott}
\end{center}
\end{figure}

The studies\cite{Haule:2009ys, Johannes:2009kx, Medici:2011cr, Quan:2012dq} have demonstrated that the Hund's rule coupling between the on-site $d$-orbital electrons plays an important role in driving the OSMott transition together with the Coulomb repulsion $U$. The Hund's rule coupling tends to align the electron spins from different orbitals into the same direction, which enhances the electron correlation and the formation of local moments.\cite{Johannes:2009kx, Wang:2010bh} \figref{fig: OSMott} displays the phase diagram of the multi-band Hubbard model obtained by the DMFT\cite{Medici:2009qf} calculation. The reader is referred to the next Chapter of this book for the detailed theoretical discussion of OSMott transitions.

\begin{figure}[htbp]
\begin{center}
\includegraphics[width=250pt]{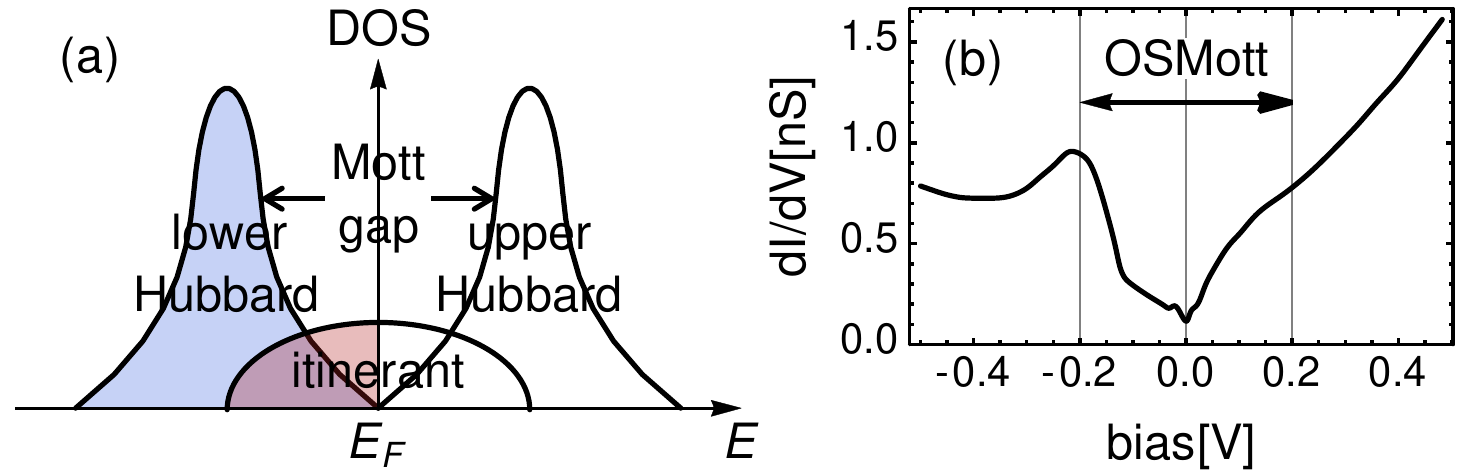}
\caption{(a) The schematic illustration of the electron density of states (DOS) in a muti-band system with the OSMott transition. Here the itinerant electron band as well as the lower and upper Hubbard bands coexist. Those electrons that fill up the lower Hubbard band form the local moments with a charge (OSMott) gap. In the hybrid model of Eq. (3), the doped electrons are assumed to all enter into the itinerant band such that the local moments remain unaffected. (b) The STM differential conductance spectrum of the NaFeAs compound, where a large OSMott-like gap ($\sim 0.4$eV) may be observed with the associated V-shape gap structure pinned at the Fermi level at different $x$'s. Cited from Ref.\,\onlinecite{Zhou:2012ia}.}
\label{fig: DOS}
\end{center}
\end{figure}

The OSMott state is characterized in the electron density of states by the coexisting itinerant band and the Hubbard bands (with the OSMott gap), as illustrated in \figref{fig: DOS}(a). The itinerant electrons contribute to a finite density of states at the Fermi energy, governed by the weakly correlated physics, while the local moments have no direct contribution at the Fermi energy,  as if there is a generalized Mott gap. On the other hand, the local moment degree of freedom will dominate the low-lying spin fluctuations in the magnetic channel. The coexistence of weakly and strongly correlated components is supported by the optical\cite{Moon:2010bs, Wang:2012ij} and STM \cite{Zhou:2012ia} experiments in the iron-based compound. In the BaFe$_2$As$_2$ compound, the optical measurement\cite{Wang:2012ij} has revealed a charge transfer gap of the energy scale 0.6eV opening up at low temperature. Similar large gap feature of $\sim 0.4$eV has been directly observed in the STM differential conductance spectrum of NaFe$_{1-x}$Co$_x$As compounds,\cite{Zhou:2012ia} as shown in \figref{fig: DOS}(b), where a V-shaped like feature associated with this large Mott-like gap is ``pinned'' at the Fermi energy, with a finite zero-bias density of states of the itinerant electrons where the smaller SDW and SC gaps are found at low temperatures at different $x$'s. With the increase of $x$, the electrons are doped into the FeAs layer which seem all entering the itinerant bands, leading to their rigid band shift. On the other hand, the high-energy V-shaped curves remains unchanged and pinned at the Fermi level,\cite{Zhou:2012ia} indicating that no significant doped charges go to the Mott localized bands, lending support to \figref{fig: DOS}(a) and the hybrid model of Eq. (3).

\section{Two-Fluid Description for Iron-Based Superconductors}

In the Introduction, we have given both experimental evidence and theoretical consideration justifying a phenomenological description of the iron-based superconductor, namely, the coexisting itinerant and localized electrons or the so-called hybrid theory. In this section, we shall present a detailed and systematic model study along this line of thinking, which provides a unified understanding for the basic physics of magnetism and superconductivity in the iron-based materials.

\subsection{Two-Fluid Description Based on the Hybrid Model}

In the hybrid theory, the minimal model of Eq. (3) is a mixture of the weakly correlated Fermi liquid physics of itinerant electrons and the strongly correlated Mott physics of local moments. Without doping into the local moment degree of freedom, the charge carriers remain in a Fermi liquid state, which is much simplified as compared to the doped $t$-$J$ type model in Eq. (2). 

The study of Eq. (3) in various parameter regimes\cite{Kou:2009bd, You:2011ud} has demonstrated that the hybrid theory is capable of explaining the SDW and SC ordered phases as well as the normal state properties in the iron-based superconductors. However, in order to accommodate the experiments of the iron-pnictide in different phases consistently, it has been further found\cite{Kou:2009bd, You:2011ud} that the zeroth-order ground state of the local moments described by $H_\text{loc}$  (i.e., without considering the interaction term $H_\text{cp}$ in Eq. (3)) should be in or very close to a short-range ordered CAMF state instead of deep in a long-range ordered CAMF state. For example, if $H_\text{loc}$ is described by a spin $S=1$ $J_1$-$J_2$ Heisenberg model, the choice of $J_1/J_2$ should be close to the so-called spin liquid regime. In other words, in the hybrid theory, the experimentally observed CAMF order is not simply associated with a magnetical order of the local moments themselves, as predicted by a pure local moment theory. Instead, it is an SDW ordering of the itinerant electrons which is driven by coupling to the fluctuating local moments \textit{perturbatively}. This novel understanding of the mangetic phase explains many detailed experimental features like the relatively large static magnetization vs. small SDW gap, etc.\cite{Kou:2009bd,You:2011ud} 

Therefore, as a phenomenological approach, we may further simplify the hybrid model by treating the itinerant electrons and local moments as two separate liquids, i.e., Fermi liquid and spin liquid. We may call it a two-fluid description, similar to the liquid Helium in which the famous two-fluid model\cite{Tisza:1938uz} was proposed to describe the coexistence of the superfluid and normal fluid components. Then the effective Hamiltonian (3) is rewritten in the following form
\begin{equation}\label{eq: H two fluid}
H=H_c+H_b+H_{cb},
\end{equation}
with 
\begin{equation}\label{eq: Hb}
\begin{split}
H_\text{it}\rightarrow H_c &,\\
H_\text{loc}\rightarrow H_b & =\sum_{i,j}(\eta_{ij}b_i i\sigma_2 b_j+\chi_{ij}b_i^\dagger b_j+h.c.)+\sum_{i}\lambda b_i^\dagger b_i,\\
H_{cp}\rightarrow H_{cb}&=-J_0\sum_{i}(c_i^\dagger\vect{\sigma}c_i)\cdot(b_i^\dagger\vect{\sigma}b_i).
\end{split}
\end{equation}
Here the local moment ($S=1$ is assumed) is fractionalized into $S=1/2$ spinons by introducing the Schwinger bosons $b_{i\sigma}$ ($\sigma=\uparrow,\downarrow$)\cite{Arovas:1988iq} : $\vect{M}_i = \frac{1}{2}b_{i\sigma}^\dagger \vect{\sigma}_{\sigma\sigma'}b_{i\sigma'}$, such that $H_\text{loc}$ in Eq. (3) becomes a four-spinon Hamiltonian, which can be further mean-field decomposed to a spinon bilinear Hamiltonian $H_{b}$ with the spinon pairing $\eta_{ij}$ and hopping $\chi_{ij}$ terms. $H_{cb}$ represents the coupling between the itinerant electron and the spinons which follows from the effective Hund's rule coupling $H_\text{cp}$ in Eq. (3), where $J_{0}\propto J_{H}$ will be treated as a tunable model parameter. 

Finally we remark that a spin liquid is usually considered to be a short-range ordered AFM ground state of a frustrated quantum spin model, which supports fractionalized spinon excitations.\cite{Wen:1991wl, Sachdev:1991eh, Read:1991oh,Chubukov:1994ce, Senthil:2004xv,Yao:2010zi} Here we use this concept in a much loose sense: $H_b$ just provides a convenient model description of a short-range ordered CAFM state. Alternatively in Refs. \onlinecite{Kou:2009bd,You:2011ud}, an effective nonlinear $\sigma$ model has been used to describe the low-lying magnetically fluctuating local moments. In fact, as the low-energy emergent degree of freedom, the local moments in the iron-based superconductors are not necessarily well quantized spins in a strong Mott regime. The charge fluctuations due a relatively small Mott gap may also lead to higher order spin interactions  (such as the ring exchange term). The double exchange interaction by coupling to the itinerant electrons may also increase the magnetic frustration and suppress the local moment ordering tendency.

\subsection{Low Energy Collective Modes}

\begin{figure}[htbp]
\begin{center}
\includegraphics[width=200pt]{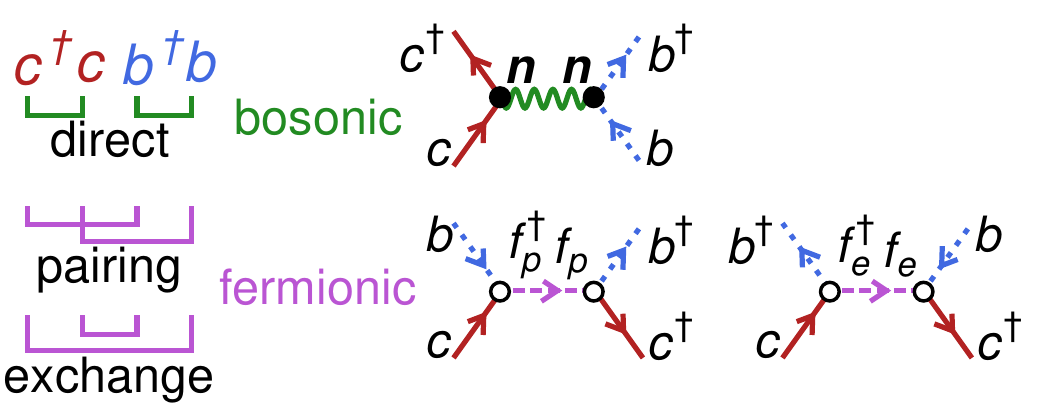}
\caption{The decomposition of the electron-spinon coupling in the direct, pairing and exchange channels. The direct channel introduces the bosonic magnon field $\vect{n}$, while the pairing and exchange channels involve the fermionic field $f_p$ and $f_e$, respectively.}
\label{fig: HS}
\end{center}
\end{figure}

In the Hamiltonian \eqnref{eq: H two fluid} of the two-fluid model, $H_{c}$ and $H_{b}$ are in bilinear forms, but $H_{cb}$ is not, which may be treated perturbatively. The four-operator Hamiltonian $H_{cb}$ has three different channels of mean-field decompositions: the direct channel, the pairing channel, and the exchange channel. 
By using the solid line $\diag{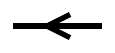}{8pt}$ to represent the itinerant electron propagator and the dotted line $\diag{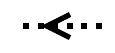}{8pt}$ for the spinon, the mean-field decomposition of $H_{cb}$ can be illustrated by \figref{fig: HS}, where different mean-field decomposition channels are mediated by different collective modes.

The direct channel involves the mean-fields $\vect{n}_{c}=\langle c^{\dagger}\vect{\sigma}c\rangle$ and $\vect{n}_{b}=\langle b^{\dagger}\vect{\sigma}b\rangle$, which represent the magnetic moments of the itinerant electron and the spinon, respectively. Here the Hund's rule coupling term is decomposed into (omiting the band and site indices)
\begin{equation}\label{eq: Hcb in n}
H_{cb}=-J_0(\vect{n}_b\cdot c^\dagger\vect{\sigma}c+\vect{n}_c\cdot b^\dagger\vect{\sigma}b - \vect{n}_b\cdot\vect{n}_c).
\end{equation}
By integrating out the itinerant electron and the spinon degrees of freedom, $\vect{n}_c$ and $\vect{n}_b$ will acquire dynamics, behaving like a collective magnon mode which describes the magnetic fluctuations. Denoting the magnon propagator by a wavy line $\diag{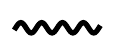}{8pt}$, then it can be depicted by the following Feynman diagrams
\begin{equation*}
\diag{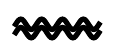}{8pt} = \diag{Gn0}{8pt}+\diag{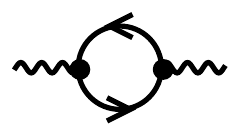}{20pt}+\diag{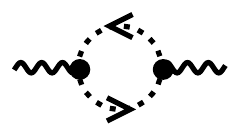}{20pt}+\diag{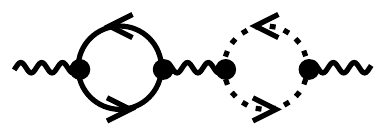}{20pt}+\cdots.
\end{equation*}

The pairing channel involves a mean-field $f_p=\langle c b\rangle$, which can be considered as a composite fermion as a bound state of the itinerant electron with the bosonic spinon. In this channel, the mean-field decomposition takes the following form
\begin{equation}
H_{cb}=-J_0(f_p^\dagger c b+h.c. - f_p^\dagger f_p).
\end{equation}
Similarly the exchange channel takes a mean-field $f_e = \langle c b^\dagger \rangle$, which can be regarded as a composite fermion bound state of the itinerant electron and anti-spinon. In this channel, the mean-field decomposition takes the following form
\begin{equation}
H_{cb}=-J_0(f_e^\dagger c b^\dagger+h.c. - f_e^\dagger f_e).
\end{equation}
By integrating out the itinerant electron and the spinon degrees of freedom, $f_p$ and $f_e$ will acquire their dynamics, which behave as composite fermions. Denoting the composite fermion propagator by a dashed line $\diag{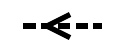}{8pt}$, the emergence of such a composite fermion can be depicted by the following Feynman diagrams
\begin{equation*}
\diag{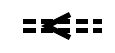}{8pt} = \diag{Gf0}{8pt}+\diag{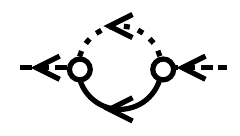}{20pt}+\diag{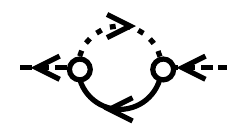}{20pt}+\diag{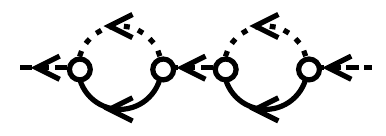}{20pt}+\cdots.
\end{equation*}
As a combination of the itinerant electron and the spinon, such composite fermions represent a unique collective charge mode in the two-fluid model whose physical consequence will be discussed later.

\subsection{Mean-Field Phase Diagram}

The SDW and the SC phases are the most prominent phases in the phase diagram of iron-based compounds. They can be understood qualitatively from the mean-field theory of the two-fluid model.

\begin{figure}[htbp]
\begin{center}
\includegraphics[width=220pt]{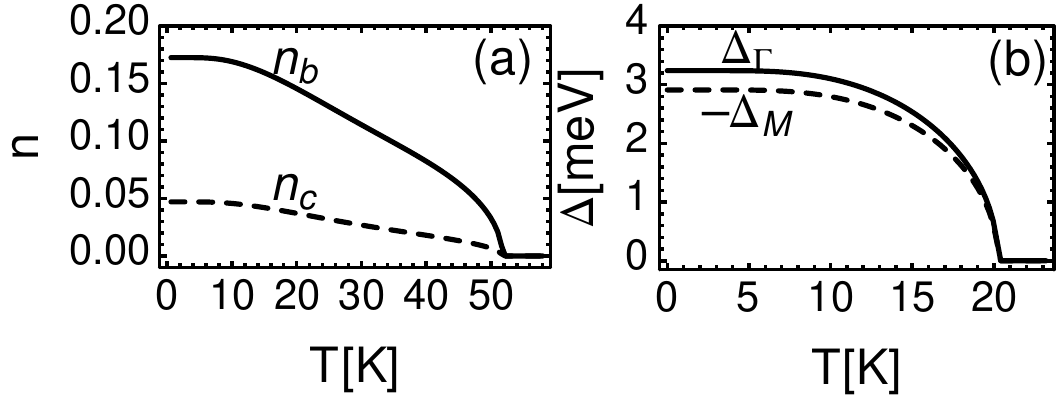}
\caption{Mean-field solutions of (a) SDW and (b) SC order parameters. $\vect{n}_c$ and $\vect{n}_b$ are the SDW/CAMF order parameters of the itinerant electron and the spinon, respectively. $\Delta_\Gamma$ and $\Delta_M$ are the SC order around the $\Gamma$ and $M$ Fermi pockets, respectively. Cited from Ref.\,\onlinecite{You:2013fk}.}
\label{fig: SDWSC}
\end{center}
\end{figure}

Let us start with the SDW phase, in which the magnetism is neither fully itinerant nor fully local origin.\cite{Johannes:2009kx}
Based on the hybrid theory,\cite{Kou:2009bd,You:2011ud} the SDW phase is the consequence of a joint effort of the coupled itinerant and localized degrees of freedom. Due to the Hund's rule coupling, the SDW ordering of the itinerant electrons, $\vect{n}_c=\langle c^\dagger \vect{\sigma} c\rangle$, provides an effective ``Zeeman-like'' field to the spinons, which polarizes the spinons along the same spin directions. In return, a spinon magnetic ordering $\vect{n}_b=\langle b^\dagger \vect{\sigma} b\rangle$ will act back on the itinerant electrons, helping to stabilize the SDW ordering. The itinerant electron and the local moment will thus mutually polarize each other, as described by the mean-field decomposition in \eqnref{eq: Hcb in n}. Such a positive feedback will lead to the simultaneous ordering of both the itinerant electron and the local moment SDW order parameters $\vect{n}_c$ and $\vect{n}_b$. Combined with the itinerant electron and the spinon band structures, the mean field solution of $\vect{n}_c$ and $\vect{n}_b$ can be determined self-consistently,\cite{You:2013fk} as shown in \figref{fig: SDWSC}(a). 

As the doping level increases, the nesting instability of the itinerant electron Fermi surface is suppressed.\cite{Kou:2009bd,You:2011ud,You:2013fk} Without the SDW ordering, the local moments will remain in a spin liquid state. The gapped para-magnon excitation in the spin liquid state will nevertheless drive a Cooper pairing instability of the itinerant electrons. Here the magnon plays the same role\cite{Scalapino:1986os} as the phonon in a conventional superconductor. Thus, in the two-fluid model, the iron-based superconductor is basically still a BCS superconductor with the charge carrieres provided by the itinerant electrons and the pairing glue by the local moment fluctuations. 

In a magnon-gluon BCS theory,\cite{You:2011ud,You:2013fk} there exists a cutoff frequency $\omega_D$ of magnons (similar to the Debey frequency of phonons). The SC transition temperature is controlled by this energy scale $\Delta\propto\omega_D$. Comparing with the phonon, the magnon cutoff frequency can be higher by one order of magnitude, which explains why the iron-based superconductor can support a relatively high $T_c$. In the singlet channel, the magnon mediates a repulsive interaction.\cite{Kou:2009bd,You:2011ud,You:2013fk} Thus the Fermi pockets at $\Gamma$ and $M$ points, connected by the magnon momentum (i.e. the magnetic ordering momentum) should take the opposite pairing sign, leading to the $s_\pm$-wave pairing symmetry similar to the itinerant theory\cite{Mazin:2009rp,Parish:2008zr,Ummarino:2009lg} \figref{fig: SDWSC}(b) shows the mean field calculation of the SC order parameters in the electron doped case.\cite{You:2013fk}

\begin{figure}[htbp]
\begin{center}
\includegraphics[width=140pt]{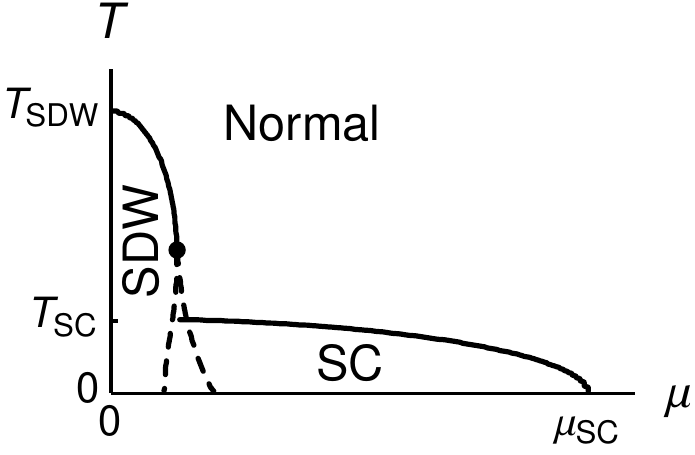}
\caption{Mean-field phase diagram. The solid (dashed) curves are 2nd (1st) order phase transition boundaries. The black dot is a tricritical point. Cited from Ref.\,\onlinecite{You:2011pi}.}
\label{fig: phase}
\end{center}
\end{figure}

Putting the SDW and SC together,  a mean-field phase diagram can be mapped out, as shown in \figref{fig: phase}. With increasing doping, the SDW transition line may end at a tricritical point that can further split into first order transitions.\cite{Chubukov:2008bu,Chubukov:2009ta,You:2011pi} The coexistence/competition of the SDW and SC order in the intermediate phase has also been discussed in Ref.\,\onlinecite{Vorontsov:2009mq,Vorontsov:2010wa}.

In summary, the overall phase diagram of the iron-based compound may be qualitatively understood in the hybrid/two-fluid picture. Here in both the SDW and the SC phases, the coexistence of the itinerant electrons and the local moments are crucial to the underlying mechanisms. In the SDW phase, the itinerant electrons and the local moments mutually interact to facilitate a joint ordering, while in the SC phase, the itinerant charge carriers are paired via the gluon provided by the local moment fluctuations.

\subsection{Spin Dynamics}
\subsubsection{NMR Knight Shift}

\begin{figure}[htbp]
\begin{center}
\includegraphics[width=140pt]{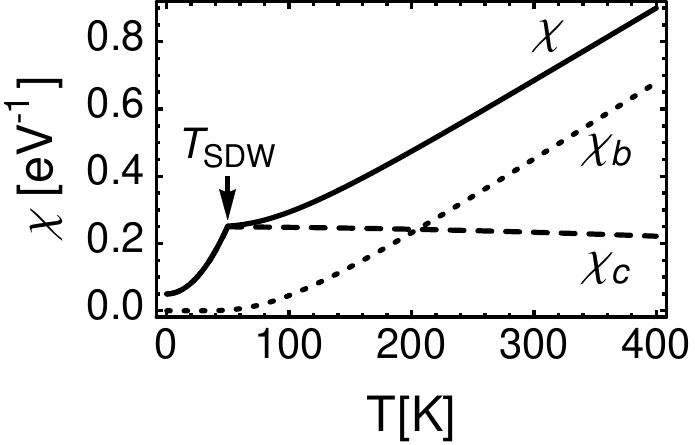}
\caption{Typical behavior of the Knight shift (uniform spin susceptibility) of the two-fluid model. $\chi_c$ and $\chi_b$ are the contributions form the Fermi liquid and the spin liquid, respectively.}
\label{fig: uniform}
\end{center}
\end{figure}

The two-fluid behavior is also reflected in the NMR Knight shift. The Knight shift basically measures the uniform spin susceptibility $\chi$ as a function of temperature, which includes the contributions from both the itinerant electron $\chi_c$ and the local moment $\chi_b$. At the RPA level, one finds 
\begin{equation}\label{eq: RPA}
\chi=\frac{\chi_c+\chi_b}{1-J_0^2\chi_c\chi_b},
\end{equation}
where $J_0$ is the effective Hund's rule coupling strength. For a Fermi liquid, $\chi_c$ follows the Pauli susceptibility behavior, which is almost independent of $T$. For a spin liquid, $\chi_b$ follows a linear-$T$ behavior as shown before.\cite{Kou:2009bd, You:2011pi}. Since $\chi_c$ and $\chi_b$ are much smaller compared to $J_0^{-1}$, the denominator in \eqnref{eq: RPA} is not important, which is simplified to $\chi\simeq \chi_c+\chi_b$. Its typical behavior is shown in \figref{fig: uniform}, together with $\chi_c$ and $\chi_b$.

At higher temperatures, the spin liquid (local moment) dominates the linear-$T$ behavior, while  the Fermi liquid behavior takes over as the temperature lowers, where the Knight shift would saturate to a constant Pauli susceptibility. However, further taking into account of the SDW transition, the Fermi liquid susceptibility will suddenly be reduced below the transition temperature $T_\text{SDW}$ (see \figref{fig: uniform}), as the Fermi surface density of states gets depleted due to the SDW gap opening. The coexistence of the high-temperature linear-$T$ behavior and the low-temperature SDW gap-opening behavior in a single curve of the Knight shift\cite{Imai:2009mw,Ma:2011ee,Michioka:2010fc,Nakai:2009xw,Ning:2009ss,Ning:2010qq} once again supports the two-fluid description.

\subsubsection{INS Spectrum}

\begin{figure}[htb]
\begin{center}
\includegraphics[width=190pt]{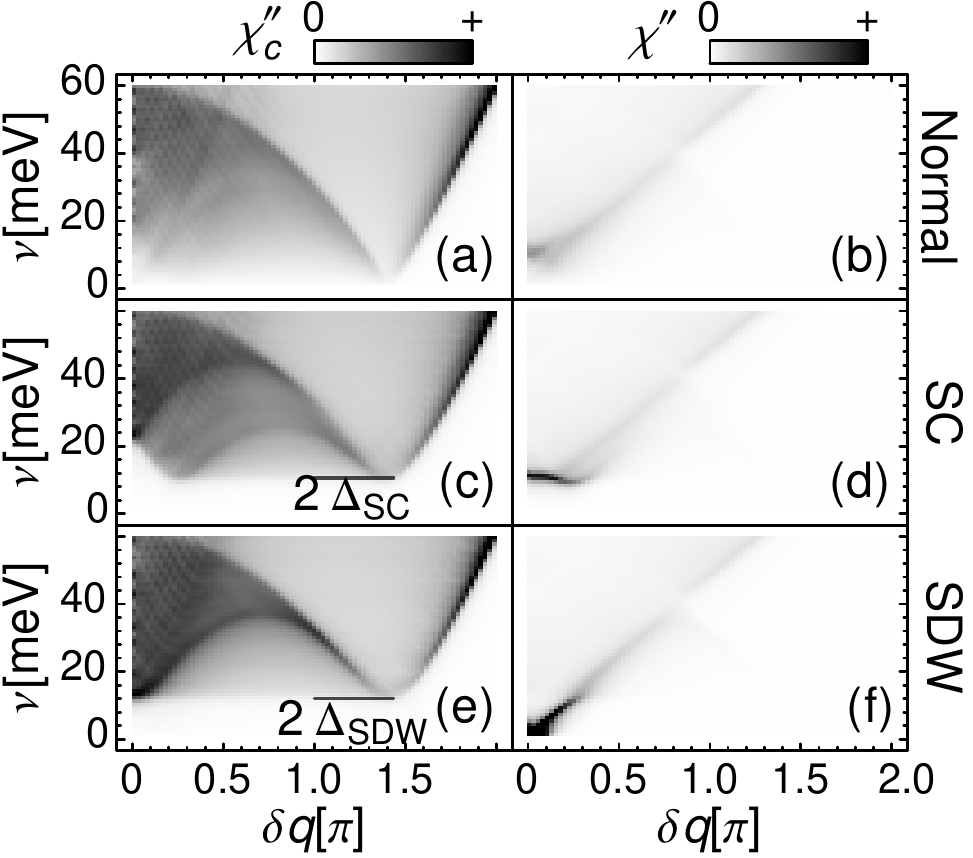}
\caption{The calculated dynamic spin spectral function of the itinerant electrons, $\chi_c''(q)$, in (a,c,e) vs. the total dynamic spectral function, $\chi''(q)$, in (b,d,f). Here (a,b) are in the normal state, (c,d) are in the SC phase, and (e,f) are in the SDW phase (for the transverse mode). The zero-momentum point is shifted to the magnetic ordering momentum, such that $\delta\vect{q}$ measures the deviation from the ordering momentum.}
\label{fig: INS}
\end{center}
\end{figure}

The two-fluid character is manifested not only in the static spin response， but also in the dynamic spin fluctuations. The INS experiment can probe the dynamic spin-spin correlation $\chi(q)$, which again includes the contributions from both the itinerant, $\chi_c(q)$, and local moment, $\chi_b(q)$, degrees of freedom at the RPA level:
\begin{equation}\label{eq: RPA_q}
\chi(q)=\frac{\chi_c(q)+\chi_b(q)}{1-J_0^2\chi_c(q)\chi_b(q)},
\end{equation}
with $q=(\nu,\vect{q})$ being the frequency-momentum vector. While the dynamic spectral function of the local moments, described by $\chi"_b(q)$, is not much affected in different phases, the contribution from the itinerant electrons, $\chi"_c(q)$, is quite sensitive to different states of the itinerant electrons, as shown in \figref{fig: INS}(a,c,e).

In the normal state, the dynamic spin susceptibility of the Fermi liquid simply forms a Stoner continuum as in \figref{fig: INS}(a). The total dynamic spin susceptibility measured by INS spectrum shows a massive dispersion relation of the local moment magnon \figref{fig: INS}(b), with the certain degree of blurring due to the self-energy correction brought by the Stoner continuum of the itinerant electron.

In the SC phase, the Stoner continuum is gapped up by the SC gap $\Delta_\text{SC}$ as in \figref{fig: INS}(c). The discontinuity of $\chi_c''(q)$ at the gap edge leads to the divergence of $\chi_c$ according to the Kramer-Kronig relation. Then from \eqnref{eq: RPA_q}, the denominator could easily vanish given enough coupling strength $J_0$, leading to the divergence of $\chi(q)$. This gives rise to the spin-resonance at the energy scale of $2\Delta_\text{SC}$ around the magnetic ordering momentum in the SC phase, as shown in \figref{fig: INS}(d), a phenomenon that has been observed\cite{Argyriou:2010vn,Chi:2009uq,Christianson:2008ky,Christianson:2009ir,Harriger:2012tg,Inosov:2010fk,Li:2009kx,Lumsden:2009ij,Lumsden:2010uq,Lynn:2009fk,Qiu:2009fl,Seo:2009ul,Taylor:2011qe,Wang:2012zr,Wen:2010zr,Xu:2011uq} in various families of iron-based superconductors. 

In the SDW phase, the Stoner continuum of the itinerant electron is gaped up by the SDW gap $\Delta_\text{SDW}$, as shown in \figref{fig: INS}(e). Due to the broken spin rotational symmetry in the SDW phase, the spin fluctuations can be divided into the transverse fluctuation (perpendicular to the magnetization direction) and the longitudinal fluctuation (parallel to the magnetization). For the transverse fluctuation, a gapless Goldstone mode will emerge inside the SDW gap shown in \figref{fig: INS}(f), as the new poles of $\chi(q)$. The gapless Goldstone mode is a consequence of the spontaneous broken spin-rotation symmetry in forming the SDW joint ordering. While for the longitudinal fluctuation, the magnon mode will remain gapped. The gap of the longitudinal mode is about $2\Delta_\text{SDW}$, which is related to the spin fluctuation asscociated with the itinerant electron, since a minimal $2\Delta_\text{SDW}$ energy is required to excite an SDW electron-hole pair as the longitudinal mode. Thus, a low-energy longitudinal spin fluctuation observed in the SDW phase is an evidence for the presence of the itinerant magnetism. 
\begin{figure}[htbp]
\begin{center}
\includegraphics[width=240pt]{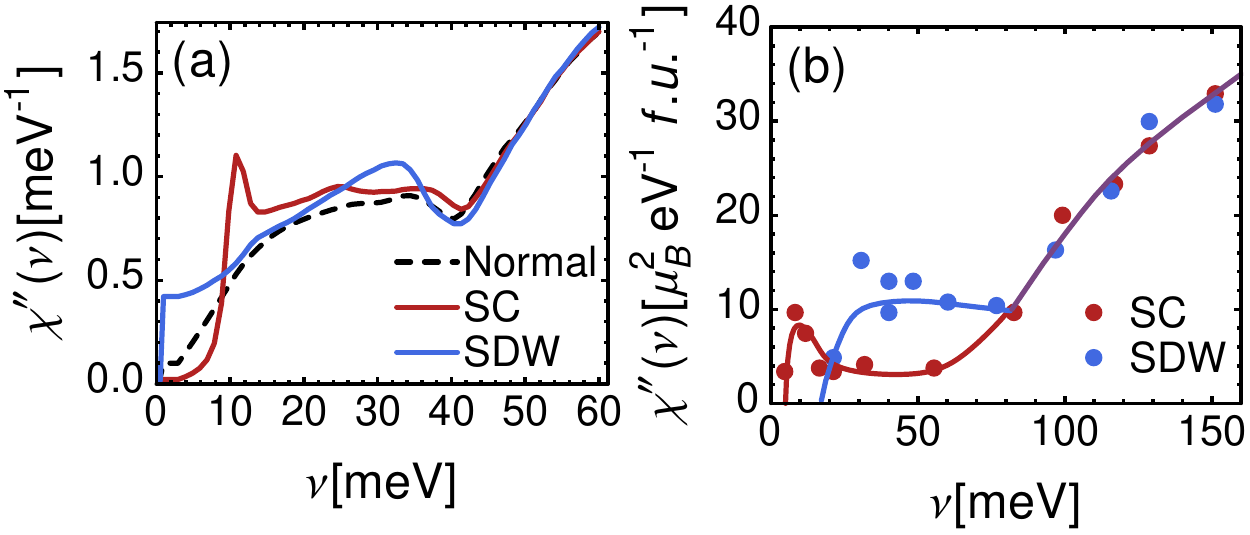}
\caption{Local (momentum-integrated) dynamic spin spectral function $\chi''(\nu)$ in different phases. (a) Theoretical calculation according to the two-fluid model, cited from Ref.\,\onlinecite{You:2013fk}. (b) INS experiment data for BaFe$_{2-x}$Ni$_x$As$_2$ compounds, cited from Ref.\,\onlinecite{Liu:2012rg}. The kink structure indicates that the upper edge of the Stoner continuum, above which the spin fluctuation spectrum is almost not affected by the itinerant electron.}
\label{fig: spin dynamics}
\end{center}
\end{figure}

Therefore, the spin fluctuations have been studied\cite{You:2013fk} as an RPA combination from the two-fluid components. The low-energy spin fluctuations are much affected by states of the itinerant electron, which behaves differently in different phases. On the other hand, the high-energy spin fluctuations mainly reflect the dynamics of the local moment, which is much less sensitive to either SDW or SC ordering. The low- and high-energy sectors are separated by the upper edge, a kink structure, of the Stoner continuum of the itinerant electron, illustrated in \figref{fig: spin dynamics}(a). Such behavior is consistent with the experiments,\cite{Argyriou:2010vn,Liu:2012rg} as shown in \figref{fig: spin dynamics}(b). Moreover, at the low energies, the RPA correction naturally gives rise to a spin-resonance mode in the SC phase as well as a Goldstone mode (spin wave excitation) in the SDW phase. 

\subsection{Charge Dynamics}

\subsubsection{Resistivity}

The transport property of the iron-based compound is determined by the itinerant electron near the Fermi surface, which basically follows the Fermi liquid behavior. The scattering of the itinerant electron with the underlying local moment fluctuation provides an important source of dissipation. The self-energy correct of the itinerant electron due to the electron-spinon scattering can be evaluated on the RPA level.\cite{You:2011ud} At low temperature $T$ and small frequency $\omega$, it was shown\cite{You:2011ud} that the imaginary part of the self-energy approximately follows the $\Im\Sigma\sim\omega^2$ or $T^2$ behavior (depending on which one is greater). This gives rise to the $\rho\sim T^2$ dependence of the resistivity at low temperature, typical for the Fermi liquid.

In the SDW phase, such a $T^2$ behavior of the resistivity is in competition with the thermal activation behavior across the SDW gap. As temperature increases, the SDW gap is suppressed, and more itinerant electrons are thermally excited across the SDW gap to contribute to the conductivity. So this activation effect tends to reduce the resistivity with the temperature, which is in opposite to the $T^2$ behavior. The competition between these two factors may eventually lead to a hump in the resistivity curve for under-doped compounds in the SDW phase, as the $x=0.04$ curve in \figref{fig: resistivity}. 

According to the Kramers-Kronig relation, the real part of the self-energy should follow the $\Re\Sigma\sim\omega$ behavior, which leads to the band renormalization effect. As has been reported in various ARPES experiments,\cite{Lu:2008uq,Liu:2008yo,Yang:2009kx,Lu:2009lp,Xia:2009dz,Liu:2009zh,Borisenko:2010qo,Cui:2012kl} all pockets are shallower than the bare band structure predicted by the DFT calculations\cite{Cao:2008bh,Singh:2008xq,Mazin:2008kn,Kuroki:2008qq,Deng:2009tg,Graser:2010qw} The mass enhancement factor can be as strong as 2 to 3.\cite{Anisimov:2009kx,Ferber:2012nx}

\subsubsection{STM Spectrum}

\begin{figure}[htbp]
\begin{center}
\includegraphics[width=200pt]{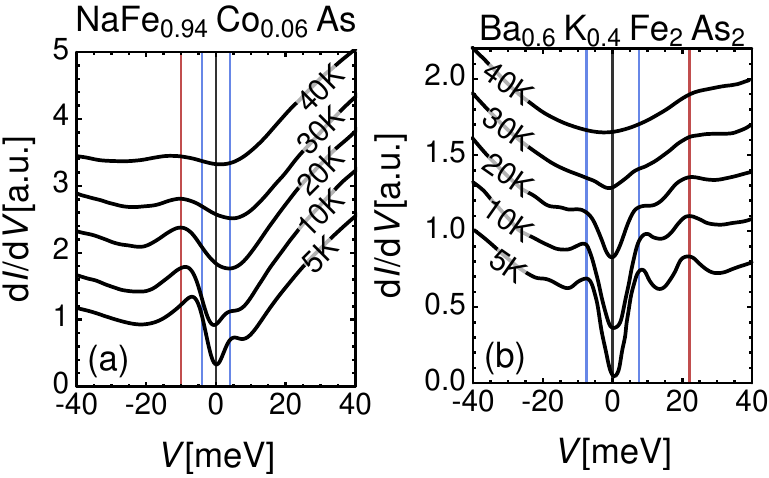}
\caption{Experimental data of STM differential conductance in (a) the electron-doped compound and (b) the hole-doped compound. The black line indicates the Fermi level. The blue lines mark out the SC coherence peak. The red line marks out the hump structure. Cited from (a) Ref.\,\onlinecite{Zhou:2012ia} and (b) Ref.\,\onlinecite{Wang:2013sh}.}
\label{fig: STM data}
\end{center}
\end{figure}

The STM differential conductance ($\mathrm{d}I/\mathrm{d}V$ spectrum) basically measures the electron local density of states. The STM studies\cite{Massee:2010tt,Zhou:2012ia,Wang:2013sh} in many iron-based superconductors have discovered a hump-dip feature in the normal phase, as shown in \figref{fig: STM data}. Starting from the low temperature SC phase and raising the temperature into the normal phase, an hump-dip feature was leftover around the Fermi level in the normal state after the closure of the SC gap (at around 20K for NaFe$_{0.94}$Co$_{0.06}$As and 40K for Ba$_{0.6}$K$_{0.4}$Fe$_2$As$_2$), see \figref{fig: STM data}. It was also found that the dip structure is locked to the Fermi surface under different doping with asymmetric line shape. Moreover, from the electron-doped \figref{fig: STM data}(a) to the hole-doped \figref{fig: STM data}(b) compounds, the $\mathrm{d}I/\mathrm{d}V$ spectrum is particle-hole reflected about the Fermi level.

Given the observed facts, this hump-dip structure can be explained neither as an SDW gap due to its locking with the Fermi level, nor as a SC gap due to the asymmetric line shape. One possible explanation is that the hump structure represents a charge resonance mode, originated from the composite fermion discussed in the above two-fluid model. Due to the effective Hund's rule interaction between the itinerant electron and the spinon, they may be bound together into a composite fermion, which carries one electron charge and an integer spin. For an intermediate coupling strength, the composite fermion mode will emerge near the Fermi surface within the spin gap. In a finite doping, the electron spectrum is particle-hole asymmetric, so is the composite fermion mode about the Fermi level. It is found\cite{You:2013fk} that the composite fermion mode will emerge from below the Fermi energy for the electron-doping, and from above the Fermi energy for the hole-doping. If we regard the hump structure in the STM spectrum as the signal of the composite fermion mode, then the asymmetric line shape and the doping dependence can both be understood consistently as below.\cite{You:2013fk} 

\begin{figure}[htbp]
\begin{center}
\includegraphics[width=120pt]{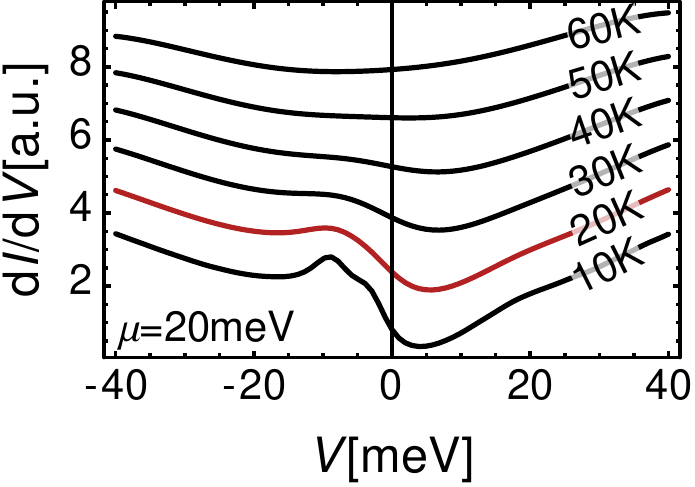}
\caption{Calculated composite fermion contribution to the STM spectrum, for the electron doped case. The asymmetric hump-dip structure is locked to the Fermi level (marked by the vertical black line), and is gradually smeared out upon raising the temperature. Cited from Ref.\,\onlinecite{You:2013fk}.}
\label{fig: dIdV}
\end{center}
\end{figure}

In \figref{fig: dIdV}, the composite fermion mode is reflected in the $\mathrm{d}I/\mathrm{d}V$ spectrum via the inelastic electron tunneling.\cite{Hahn:2000fk} If the electron from the STM tip tunnels into the sample with an energy higher than that of the composite fermion, new tunneling channels will be opened up, leading to the hump structure in the $\mathrm{d}I/\mathrm{d}V$ spectrum. The hump appears at the energy scale of the composite fermion mode, which is locked to the Fermi level by the spin gap. The calculation is done for the electron-doped case. For the hole-doped case, the spectrum will simply reversed with respect to the Fermi level, which is consistent with the experimental observations.

\section{Summary}

\begin{figure}[htb]
\begin{center}
\includegraphics[width=240pt]{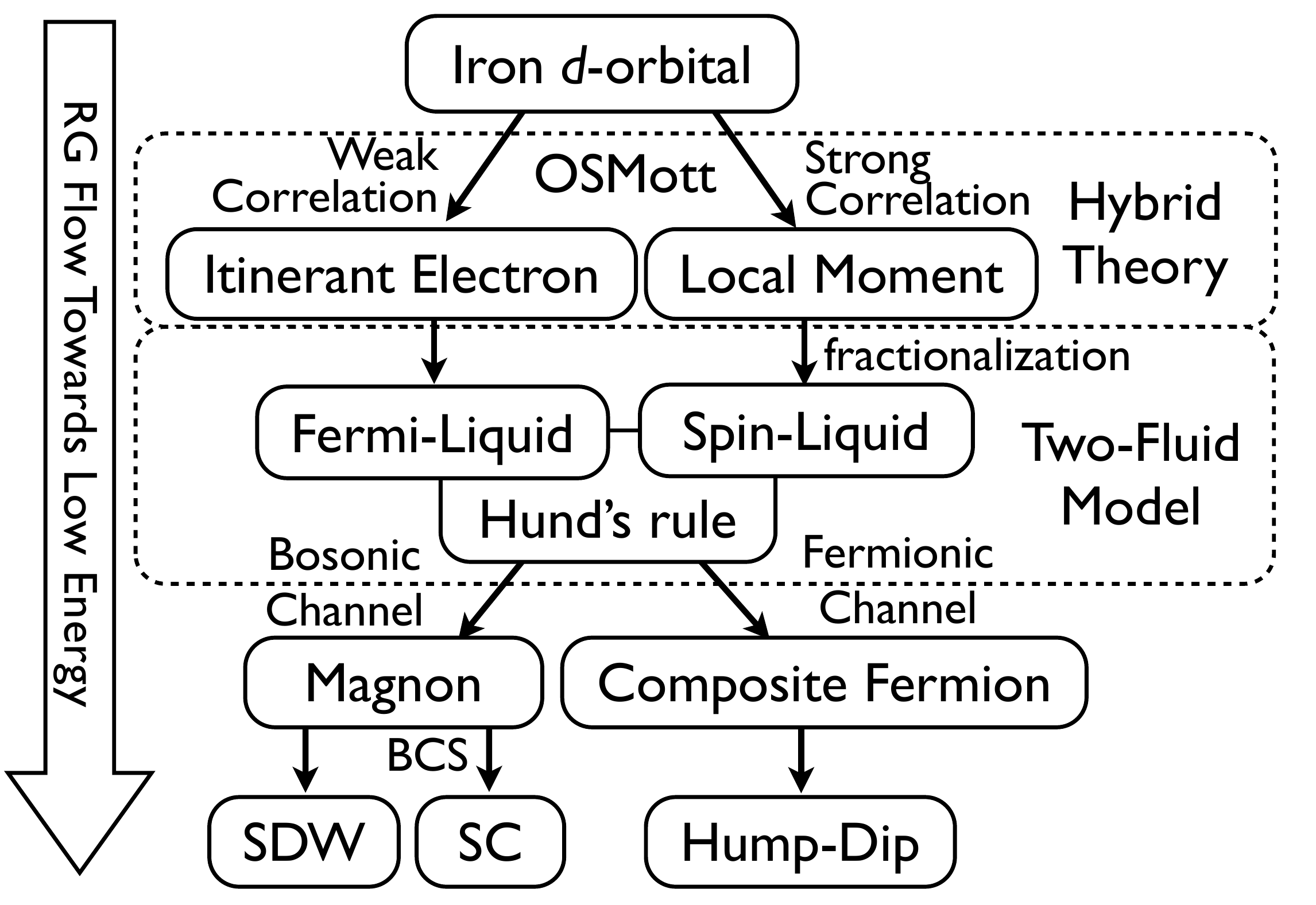}
\caption{Relevant degrees of freedoms and the effective theories emerging with the RG flow towards low energies are schematically illustrated. Here the hybrid theory provides a minimal, phenomenological description underlain by an OSMott transition. The SDW and SC states, together with the normal state of the iron-based superconductors, may be understood by a unified framework of a two-fluid model in the hybrid theory.}
\label{fig: RG flow}
\end{center}
\end{figure}

In the present Chapter, we presented a minimal, phenomenological description of the low-energy physics in the iron-based superconductors. The general framework and physical consequences are summarized in \figref{fig: RG flow} from the viewpoint of renormalization group.

Strarting from the iron $3d$ orbital electrons, their microscopic dynamics may be described by a multi-band Hubbard model \eqnref{eq: Hubbard}. Due to the multi-band and intermediate correlation characters, an orbital-selective Mott transition becomes possible.  Under an RG flow, the difference among different electron bands may be amplified, which then leads to distinct RG fixed points. Some bands flow to the strong correlation fixed point of the local moments, while the others flow to the weak correlation fixed point of the itinerant electrons. The itinerant electrons and the local moments can thus coexist in the system in the RG sense, coupled together via a residual Hund's rule interaction. This constitutes the basic rationale for the hybrid theory in Eq. (3).

At a lower temperature, the itinerant electrons form a Fermi liquid, characterized by well-defined quasi-particles around the Fermi pockets, which can further experience typical Fermi-liquid instabilities such as SDW and SC. On the other hand, the local moments remain disordered due to strong quantum fluctuations, which may be modeled by a spin liquid with fractionalized bosonic spinons by a two-fluid model [Eq. (4)]. 

In the two-fluid model, the Fermi liquid of the itinerant electron and the  spin liquid of the local moment are coupled together by a residual Hund's rule coupling, which may be treated perturbatively in a weak or intermediate coupling strength. There are two types of low-lying collective modes arising from this coupled two-fluid model.

In the bosonic channel, a magnon-like excitation as a bound state of the spinons reflects the strong magnetic correlations of the local moments. It couples to the itinerant electrons to induce an SDW ordering of the latter, while simulaneously lead to a CAMF ordering of the local moments in the magnetic phase of the iron-based superconductor. On the other hand,  the magnon-mediated effective pairing between the itinerant electrons competes with the SDW ordering near the Fermi surfaces, resulting in an SC state in proper parameter regimes.

A unique prediction of the two-fluid model is that besides the usual quasiparticle excitation, a new composite fermion mode may emerge as a bound state of the itinerant electron and the local moment under an intermediate Hund's rule interaction. It carries the same charge as an itinerant electron but with a different spin quantum number, which participates in the low energy charge transport and leads to the hump-dip structure observed in the STM inelastic electron tunneling spectrum (IETS) in the iron-based compounds.

Finally, we point out that the iron-based superconductor is not the only known physics system that may possess the coexisting itinerant electrons and local moments. The heavy fermion system\cite{Andres:1975mn, Steglich:1979ph} discovered in the 1970's is already one of such examples. The theoretical framework \cite{Nakatsuji:2004we,Yang:2008hc,Yang:2008bs} at low energy also contains two fluid components: the Fermi liquid of the coherent electrons and the Kondo lattice, in which the local moment at each site couples to the itinerant electron via the antiferromagnetic Kondo interaction. Such Kondo lattice model looks similar to the hybrid model of the iron-based superconductor. One of main distinctions lies in whether the coupling described by the $H_{cp}$ term is antiferromagnetic (Kondo) or ferromagnetic (Hund's rule). But this difference is important. The RG equation $\mathrm{d}J/\mathrm{d}\ln\Lambda=-2\rho J^2$ is sensitive to the sign of the coupling $J$: the Kondo coupling can flow to infinity towards low energy, while the Hund's rule coupling flows to zero. So in some sense, the iron-based superconductor system is simpler comparing to the heavy fermion system because the Hund's rule coupling can be treated perturbatively. Even in the single-band $t$-$J$ model, via the fractionalization, a two-fluid description of the spin correlations has been recently proposed in the superconducting state, where the Mott localized spins and doping-induced hopping effect are described by a two-component RVB structure in the ground state wave function.

\begin{acknowledgments}
Previous collaborations and discussions with S.P. Kou, F. Yang, Y.Y. Wang, and T. Li related to the present work are acknowledged. This work was supported by the NBRPC grant no. 2010CB923003.
\end{acknowledgments}
\bibliography{refs}
\bibliographystyle{apsrev}
\end{document}